\documentclass[grl,final]{agutex}

\usepackage{amsmath, amssymb}
\usepackage{array}
 \usepackage{graphicx}
 \setkeys{Gin}{draft=false}
\authorrunninghead{SUN, TEUNISSEN, EBERT}

\titlerunninghead{STREAMER FORMATION}

\authoraddr{U. Ebert, CWI,
P.O. Box 94079, 1090 GB Amsterdam, The Netherlands
(U. Ebert@cwi.nl).}

\authoraddr{A. B. Sun, CWI,
P.O. Box 94079, 1090 GB Amsterdam, The Netherlands
(a.sun@cwi.nl).}

\authoraddr{J. Teunissen, CWI,
P.O. Box 94079, 1090 GB Amsterdam, The Netherlands
(H.J.Teunissen@cwi.nl).}

\begin{document}

%% ------------------------------------------------------------------------ %%
%
%  TITLE
%
%% ------------------------------------------------------------------------ %%

\title{Why isolated streamer discharges hardly exist above the breakdown field in atmospheric air}
%
% e.g., \title{Terrestrial ring current:
% Origin, formation, and decay $\alpha\beta\Gamma\Delta$}
%

%% ------------------------------------------------------------------------ %%
%
%  AUTHORS AND AFFILIATIONS
%
%% ------------------------------------------------------------------------ %%

%Use \author{\altaffilmark{}} and \altaffiltext{}

% \altaffilmark will produce footnote;
% matching \altaffiltext will appear at bottom of page.

\authors{A.B. Sun,\altaffilmark{1}
J. Teunissen,\altaffilmark{1} and U. Ebert\altaffilmark{1,2}}

\altaffiltext{1}{Center for Mathematics and Computer Science (CWI),
P.O. Box 94079, 1090 GB Amsterdam, The Netherlands.}

\altaffiltext{2}{Department of Applied Physics,
Eindhoven University of Technology, %P.O. Box 513, 5600 MB Eindhoven,
The Netherlands.}

%% ------------------------------------------------------------------------ %%
%
%  ABSTRACT
%
%% ------------------------------------------------------------------------ %%

% >> Do NOT include any \begin...\end commands within
% >> the body of the abstract.

\begin{abstract}
We investigate streamer formation in the troposphere, in electric fields above the breakdown threshold.
With fully three-dimensional particle simulations, we study the combined effect of natural background ionization and of photoionization on the discharge morphology.
In previous investigations based on deterministic fluid models without background ionization, so-called double-headed streamers emerged.
But in our improved model, many electron avalanches start to grow at different locations.
Eventually the avalanches collectively screen the electric field in the interior of the discharge.
This happens after what we call the `ionization screening time', for which we give an analytical estimate.
As this time is comparable to the streamer formation time, we conclude that isolated streamers are unlikely to exist in fields well above breakdown in atmospheric air.
\textbf{Citation:} Sun, A. B., J., Teunissen, U., Ebert (2013), Why isolated streamer discharges hardly exist above the breakdown field in atmospheric air, {\it Geophys. Res. Lett.,} 40, 2417-2422, doi: 10.1002/grl.50457.
\end{abstract}

%% ------------------------------------------------------------------------ %%
%
%%  BEGIN ARTICLE
%
%% ------------------------------------------------------------------------ %%

% The body of the article must start with a \begin{article} command
%
% \end{article} must follow the references section, before the figures
%  and tables.

\begin{article}

%% ------------------------------------------------------------------------ %%
%
%  TEXT
%
%% ------------------------------------------------------------------------ %%

\section{Introduction}
\label{introduction}
Streamers play a key role in the early stages of atmospheric discharges;
they appear, e.g., in lightning inception, in the streamer coronas of lightning leaders and of jets, and in sprite discharges. %~\citep{ebert10}.
The late D.D. Sentman liked to call streamers the ``elementary particles'' of discharge physics.

Streamers are rapidly growing plasma filaments that penetrate into non-ionized regions due to the electric field enhancement at their tips.
When the local electric field exceeds the breakdown threshold of a gas, the neutral gas molecules start to become ionized by impact of electrons with energies above 12 eV.
While the ionization density grows, charged particles move in the electric field and form space charge regions that modify the field.
The ionization then grows rapidly at channel edges where the field is enhanced, while the electric field is suppressed in the ionized interior.
In this manner long ionized channels, so-called streamers, can grow.
Positive or negative streamer channel heads have to be distinguished depending on the net charge in their heads;
  they propagate along or against the direction of the electric field.

We present a new view on streamer formation in fields above the breakdown threshold.
Recently,~\citet{luquevazquez12} have shown the importance of detachment from negative ions for delayed sprite formation in the mesosphere.
Here, we show that this mechanism also changes our understanding of streamer discharges in the troposphere.

In the past 30 years, simulations that model electrons and ions as densities have developed into a key method for exploring streamer physics.
Most simulations are effectively performed in two dimensions (2D), using a longitudinal and a radial coordinate, hence assuming cylindrical symmetry of the streamer.
The emergence of a double-headed streamer, with a positive and a negative growing end, was first seen in simulations by~\citet{dhali85}.
The nonlocal photoionization mechanism that allows positive streamers to propagate in air, was first implemented by~\citet{Kulikovsky};
he also extrapolated his numerical results and suggested that such streamers grow exponentially in fields above the breakdown value.
Similar observations were later made by~\citet{liu04} who studied how these results depend on atmospheric altitude or on air density.
The exponentially growing single streamers in high fields also play a role in a recent theory on terrestrial gamma-ray flashes by \citet{celestin}.
\citet{Chanrion08} developed a 2D axi-symmetric PIC-MCC model to study streamers,
and found that a double-headed streamer forms at 10 km altitude, with similar initial conditions as~\citet{liu04}.
At sprite altitudes around 70~km, double-headed streamers were simulated by~\citet{liu04}, \citet{Qin12} and~\citet{Chanrion08,Chanrion}.
Most of these simulations were performed with fluid models in 2D, enforcing cylindrical symmetry.
% Fluid models do not contain natural fluctuations due to the discreteness of electrons, such as the stochastic presence of an electron in a weakly ionized region,
%     or the stochastic detachment of a single electron from a negative ion.

In the present paper, we reinvestigate streamer formation in electric fields above the breakdown value.
Such `overvolted regions' can for example form around the tip of a lightning leader.
We here assume that the field quickly rises to a value above the breakdown threshold and that it is initially homogeneous.
Although not directly corresponding to a particular physical situation, this keeps the analysis more simple and general, and it can serve as a local approximation.
% And in a smaller volume not too close to the field generating structure (e.g., a leader tip), homogeneity is a reasonable approximation.
Our findings are very different from those of the authors cited above, because our model contains essential additional features:
First, we include electron detachment from negative ions, which are present due to natural background ionization.
Second, we are able to perform our simulations in full three spatial dimensions. %, while using adaptive grid refinement to resolve the steep gradients emerging in the system.
Third, we work with a particle model, following the stochastic motion of individual electrons rather than approximating them as densities with completely deterministic dynamics.
In this manner, we include physically realistic stochastic fluctuations, in particular, in the regions with low ionization,
    similarly as \citet{Chanrion08,Chanrion}, \citet{li11,li12}, and \citet{luque11}.
The calculations are performed in atmospheric air at 1 bar.
Our results show that in a field above breakdown in air, isolated streamers are unlikely to form.
This is consistent with lab experiments:
\citet{Nijdam} and \citet{briels} observed `inception clouds' that form around electrodes when a high voltage is suddenly applied to air.
These clouds form essentially in the region where the field is above the breakdown value, and streamers only form beyond this region.
We conclude that under normal atmospheric conditions, isolated streamers hardly exist in fields well above the breakdown threshold.

\section{Model}
\label{model}
A 3D particle-in-cell code with a Monte Carlo collision scheme has been developed to simulate the dynamics of streamer formation.
In the model, electrons are tracked as particles.
Ions are immobile, as they would not move significantly on the time scales we consider.
Neutral molecules are not simulated, but they provide a background density that the electrons randomly collide with.
We include elastic, inelastic, ionizing and attaching collisions.
These collisions were implemented in the same way as in~\citet{li12}, with the same cross sections for collisions.
Photoionization is an important process in many discharges, where excited $\mathrm{N}_2$ molecules emit photons that ionize $\mathrm{O}_2$ molecules.
We use a stochastic version of the photoionization model of~\citet{Zhelezniak}, as was done before by~\citet{Chanrion08}.
Below we present the most important new features of our model.

\subsection{Natural background ionization and electron detachment}
In atmospheric air near ground pressure, background ionization is mostly present in the form of $\mathrm{O}_2^{-}$ and positive ions.
The number of free electrons is much smaller, because they quickly attach to $\mathrm{O}_2$ molecules to form $\mathrm{O}_2^{-}$.
In enclosed areas such as buildings, typical background ion densities are $10^3$ - $10^4 \mathrm{cm}^{-3}$, mostly due to the decay of radon~\citep{pancheshnyi}.
As altitude increases, cosmic radiation becomes the dominant source of background ionization~\citep{ermakov}.
\textit{Ermakov et al.} measured the concentration of negative ions in the lower atmosphere.
The ion concentration increases as altitude increases.
A level of approximately $10^3 \mathrm{cm}^{-3}$ was recorded at 5 km altitude, in agreement with estimates by \citet{hulburt} and \citet{usokin}.
Background ionization can also be present due to previous discharges~\citep{luquevazquez12, Nijdam, bourdon2010}.

Electron detachment can occur when an $\mathrm{O}_2^{-}$ ion collides with a neutral gas particle.
The probability of electron detachment from $\mathrm{O}_2^{-}$ depends on the local electric field and on the gas density.
We include electron detachment from negative ions in the model, with rate coefficients from~\citet{Kossyi}.

We remark that at mesospheric altitude, most negative background ions are $\mathrm{O}^{-}$ ions as they form by dissociative attachment at low air density.
These ions are also a source of electrons by detachment~\citep{vazquezluque2010, luquevazquez12, liu12}.

Electron storage in the form of negative ions, from which they can later be detached, combined with the strong non-local effect of photoionization
  distinguishes discharges in air from those in other gases, e.g. high purity nitrogen.
  
\subsection{Numerical techniques}
An \emph{adaptive particle management} algorithm is used to control the number of simulation particles in the code.
We use relatively more simulation particles around the streamer head, and relatively few in the streamer interior.
And where the electron density is low, electrons are tracked individually.
Details of the particle management method are given by \citet{teunissen}.

To be able to simulate larger systems, an \emph{adaptive mesh refinement} (AMR) technique is used.
The AMR method is similar to the methods of~\citet{montijn06} and~\citet{luque10,luquereview12}, but now in 3D.
The code is electrostatic, as the velocities are much smaller than the speed of light and the induced magnetic fields are negligible compared to the electric fields.
At every timestep, the electric potential is computed from the charge density by solving the Poisson equation with Fishpack~\citep{adams}.
The electric field is then the numerical gradient of the electrical potential.
To increase the performance and the maximum number of simulation particles, the particle code was parallelized using MPI (Message Passing Interface).

\section{Results and discussion}
\label{result}
We perform simulations in a gas mixture of 80\% $\mathrm{N}_2$ and 20\% $\mathrm{O}_2$, at 1 bar and 293 Kelvin.
The simulation domain is cubic, of size $(4\ \mathrm{mm})^3$.
An external electric field of 7 MV/m is applied in the negative z-direction, which is about $2.3$ times of the breakdown field $E_k$.
One electron-ion pair is placed at the center of the domain.
We first show `unrealistic' results with photoionization only, followed by `realistic' results where natural background ionization is included.
Then we indicate how these results depend on the initial presence of free electrons, and we introduce the concept of the `ionization screening time'.
Finally, we discuss discharges at higher altitudes in the atmosphere.

\subsection{Photoionization only}
We first present results with photoionization only, and no background ionization.
This is not very realistic, as some background ionization will always be present in air.
But these results help to clearly illustrate the effects of background ionization later on.
We remark that other authors have often presented results with photoionization only.

Figure 1 shows the evolution of the electron density and the electric field in three stages, from 2.67 ns to 3.12 ns.
The initial electrons are accelerated rapidly in the external electric field.
They collide with molecules and ionize them, so the number of electrons and ions increases rapidly.
Since the charged particles drift in the electric field, a negative charge layer forms at the upper tip, and a positive charge layer at the lower tip.
When space charge effects become significant, the discharge is in the streamer regime.
The positive front requires a source of electrons ahead of it to propagate.
Because these electrons have to be created by photoionization, there is a delay in the propagation of the positive side of the streamer.

After $\sim 2.7$ ns, a double-headed streamer starts to form.
The electric field at the streamer tips is approximately three times the breakdown field.
Meanwhile, new avalanches start to appear around the main streamer that formed by the initial seed in the middle.
The new avalanches are triggered by photoionization.
As the avalanches develop, they overlap and interact with the main streamer, see the second and third columns of Figure 1.
Eventually, the middle streamer is completely surrounded by new avalanches.

Similar results were presented by~\citet{li11,li12}, who used a hybrid model, a higher background field of 10 MV/m and a larger ionization seed.
Therefore, double-headed streamers form earlier in their simulations.
We also performed simulations with a background field of 5 MV/m and with all other conditions as for Figure 1.
Similar phenomena were observed as in Figure 1, but after a longer time of $\sim 8$ ns.

We notice a remarkable difference when we compare our results with 2D fluid model simulations~\citep{luque08, liu04, celestin}.
In contrast to our particle model or to the hybrid model by~\citet{li12}, or to the stochastic fluid model by~\citet{luque11}, normal fluid models cannot reveal such pronounced multi-avalanche structures in overvolted gaps.

Photo-ionization plays an essential role for positive streamer formation and propagation, if background ionization can be neglected.
Without photoionization or background ionization, only negative streamers are able to form, because there are no seed electrons for the positive streamer to grow.
This can for example be seen in simulations by~\citet{li12} and by~\citet{Chanrion}.

Because the gap is overvolted, the photo-electrons can create new avalanches in the whole space.
In an undervolted gap, photoionization would only create avalanches in regions where the electric field is enhanced, close to the streamer.
Then a pronounced streamer can emerge, with a larger radius and smoother gradients than without photoionization~\citep{Wormeester}.

\subsection{Background ionization and photoionization}
\label{sec:backgroundandphoto}
We now turn to the more realistic case where natural background ionization is included.
This important mechanism was missing in previous discharge models in air.
The initial conditions now include a homogeneous density of $\mathrm{O}_2^{-}$ and positive ions, both $10^{3} \mathrm{cm}^{-3}$.
All other conditions are the same as for the case with photoionization only.
Figure 2 shows the electron density and the electric field at 2.67 ns and 2.97 ns.
We now compare Figure 2 with the first and the second columns of Figure 1.
With background ionization, there are more new avalanches, as they can start from detached electrons as well as from photo-electrons.
Figure 1 shows that photo-electrons are mostly generated close to the discharge, within 1 mm distance.
On the other hand, detachment can happen anywhere, even though it happens faster in higher electric fields.
Therefore, the avalanches are much more distributed over the whole domain in Figure 2.
As the avalanches grow, they overlap more and more, and it is no longer possible to discern a single streamer.
Since the avalanches are close together, the electric field enhancement at their tips is reduced.

Now the difference with the results of 2D fluid model simulations is even greater.
Instead of a double-headed streamer, we see a discharge that spreads out over the whole domain.
Similar discharges were observed in laboratory experiments by~\citet{briels} and by~\citet{Nijdam}.
Around a needle shaped high voltage electrode, the field is above breakdown and an ionized `inception cloud' forms.
Farther away from the electrode where the instantaneous field drops below breakdown, the cloud destabilizes into streamer channels.

Therefore the existence of well separated accelerating streamers in the overvolted region near lightning leaders in air, as postulated by \citet{liu04} and \citet{celestin}, is unlikely.

\subsection{Dependence on the initial seed}
Overvolted gaps are sensitive to the initial conditions, because homogeneous breakdown competes with streamer-like breakdown.
All fluid model simulations referenced in this paper used big initial electron seeds, without much discussion where these electrons would come from.

For the results presented above, a single electron-ion pair was initially present in the domain.
We have also performed the simulation of section~\ref{sec:backgroundandphoto} without that initial electron.
The electron density at 2.67 ns and 2.97 ns is shown in Figure~\ref{fig:fromOnlyBgion}.
We can see that the discharge starts a bit later, due to the delay in the detachment process, and it is also more uniform.
Furthermore, we have performed simulations that start with 10 or 100 electron-ion pairs.
As expected, with more free seed electrons, the discharge initially grows faster, and is more concentrated around the initial seed.

\subsection{Ionization screening time}
\label{sec:shieldingtime}
The simulation results we have presented show only the first few nanoseconds of a discharge.
Here we will discuss what happens at later times.

If in some region the electric field suddenly rises above the breakdown threshold, then the number of free electrons will grow due to impact ionization.
The electrons drift in the field and leave positive ions behind, and this charge separation reduces the electric field in the interior.
After some time $\tau_\mathrm{is}$, the electric field in the interior drops below the breakdown threshold.
This we call the `ionization screening time'.
We note that~\citet{celestin} introduced a similar time scale, which was named `critical time'.
For screening to happen, there have to be some free electrons in the overvolted region.
These are clearly present above $\sim 60$ km, but in the troposphere they can appear, for example, due to electron detachment from $\mathrm{O}_2^{-}$ ions.

We first determine $\tau_\mathrm{is}$ using a plasma fluid model, then we give a more general analytical approximation.
We use a simple geometry: there is a uniform electric field $E_0$, pointing in the negative $z$-direction,
    and the initial electron and ion density are $n_0$ for $z_0 < z < z_1$, elsewhere they are zero.
The length $z_1 - z_0$  is taken sufficiently large, then the results do not depend on this length.
Figure~\ref{fig:shielding_time} shows the ionization screening time for different fields $E_0$,
    starting from an initial density $n_0 = 10^3\;\mathrm{cm^{-3}}$ of electrons or $\mathrm{O}_2^{-}$ ions.
    
Analytical approximations to these curves are also shown, these are based on a few assumptions:
    there is no diffusion and the electrons keep their initial drift velocity $v_d(E_0)$ and effective ionization coefficient $\alpha(E_0)$.
In the geometry described above, there are then no electrons below $z_0 + v_d t$, as they drift up.
The ion density between $z_0$ and $z_0 + v_d t$ is equal to $n_0 e^{\alpha (z-z_0)}$, so the integrated charge along the $z$-coordinate is
    $(e^{\alpha v_d t}-1) e n_0 /\alpha$, where $e$ is the elementary charge.
Equating this to the charge $\epsilon_0 E_0$ needed to screen an electric field $E_0$, and solving for $t$ gives the ionization screening time
\begin{equation}
 \tau_\mathrm{is} \approx \ln\Big(1+\frac{\alpha \epsilon_0 E_0}{e n_0}\Big)/(\alpha v_d),
 \label{eq:tauE}
\end{equation}
where $\epsilon_0$ is the vacuum permittivity.
Using the values $\alpha$ and $v_d$ for the initial field $E_0$ underestimates the ionization screening time;
  to compensate for this we compute the time to shield the electric field completely to zero.
Note that in the limit $\alpha \to 0$, \eqref{eq:tauE} reduces to the dielectric relaxation time $\epsilon_0/(e n_0 \mu_0)$,
    with $\mu_0=v_d/E_0$, also known as the `Maxwell time'~\citep{pasko1998}.
If we start with negative ions, the delay due to the detachment time $\tau_D$ can be included by adding a term $\ln(1+\alpha v_d \tau_D)/(\alpha v_d)$ to~\eqref{eq:tauE}.

Figure~\ref{fig:shielding_time} also includes the detachment time~\citep{Kossyi} and the typical streamer formation time based on the Raether-Meek criterion.
When the electric field is sufficiently above breakdown, the ionization screening time is close to the streamer formation time.
Then, from these time scales alone, we can say that the presence of natural background ionization inhibits the formation of isolated streamers.
The reasoning behind this statement is as follows:
When there are many seeds, many streamers try to form.
Their collective charge separation quickly screens the electric field in the interior of the discharge, which halts the growth of streamers there.
Then the discharge grows only at the boundary of the screened, originally overvolted, region.

% After the ionization screening time, the rapid discharge development in the interior region halts.
% An isolated streamer would have to get out of the interior region within the ionization screening time.
% But, as shown below, the screening time is only slightly longer than the typical streamer formation time, so this is unlikely.
Under certain conditions, for example when the electric field rises more slowly to a value above breakdown, many streamer-like channels might form that together shield the electric field.
We leave this for future research, and note that in such a case one cannot speak of isolated streamers.

In a field of 7 MV/m we find that $\tau_\mathrm{is} = 3.2$ ns if an initial density of $10^3\;\mathrm{cm^{-3}}$ $\mathrm{O}_2^{-}$ ions is present.
These conditions correspond to the simulations shown in Figure~\ref{fig:bgion} and~\ref{fig:fromOnlyBgion}, which end at 2.97 ns.
It was not possible to simulate up to the screening time, because the number of free electrons increases rapidly before screening, dramatically slowing down our particle code.

\subsection{Discharges at higher altitudes in the atmosphere}
At higher altitudes in the atmosphere, the role of background ionization is qualitatively similar, as was stated in~\citep{Qin11}.
But there are quantitative differences:
First, based on scaling laws, the ionization density, the spatial extension and duration and the electric fields in the streamer tip scale with air density,
  but natural density fluctuations, photo-ionization and air heating do not simply scale~\citep{ebert10}.
In the mesosphere where sprite discharges occur, photoionization is about 30 times more efficient than at ground level,
  because there is no collisional quenching of the photo-emitting states.
Furthermore, cosmic radiation supplies a higher level of background ionization, also in the form of free electrons;
  therefore in the ionosphere electrons start avalanches and screening ionization waves as soon as the electric field increases;
  they are seen as halos~\citep{luque09,luquevazquez12}.
At lower altitudes like the night time mesosphere, electrons are predominantly attached, but bound as O$^-$ rather than as O$_2^-$ as at ground altitude.
Electron detachment from O$^-$ was included into discharge models by~\citet{luquevazquez12} and by~\citet{liu12}.
If previous discharges or cosmic radiation have supplied sufficient O$^-$,
    this ion density can even detach so many electrons that the local breakdown field almost vanishes~\citep{luquevazquez12}.

\section{Conclusion}
We have studied steamer formation in atmospheric air at ground altitude with a 3D particle code, including the effects of background ionization.
Due to detachment of electrons from $\mathrm{O}_2^{-}$ ions, isolated streamers do not emerge in our simulations in fields above breakdown.
Instead, many new avalanches appear, that overlap as they grow.
This creates a discharge in the whole region above the breakdown field, in agreement with experimental observations~\citep{Nijdam, briels}.
An analysis of the ionization screening time, after which there is global breakdown, leads to the same conclusion.
Photo-ionization has a similar effect as background ionization, as was already observed by \citet{li12} and \citet{luque11}.
But because photo-electrons are mostly produced close to the discharge, a more localized structure emerges.

Discharges at higher altitudes like halos and sprites evolve in a qualitatively similar manner though ionization rates due to cosmic radiation
    and reactions of electron attachment and detachment differ quantitatively.

This is the reason why double-headed streamers in the troposphere and double-headed sprites in the mesosphere rarely exist,
    as was observed by~\citet{Stenbaek-Nielsen}.
If the electric field is above breakdown in a larger region, the breakdown is rather uniform due to background ionization and electron detachment,
    while if the field is below breakdown, positive streamers emerge and propagate much more easily than negative ones~\citep{luque08,liuetal12}.

%%% End of body of article:

%%%%%%%%%%%%%%%%%%%%%%%%%%%%%%%%
%% Optional Appendix goes here
%
% \appendix resets counters and redefines section heads
% but doesn't print anything.
% After typing  \appendix
%
% \section{Here Is Appendix Title}
% will show
% Appendix A: Here Is Appendix Title
%
%%%%%%%%%%%%%%%%%%%%%%%%%%%%%%%%%%%%%%%%%%%%%%%%%%%%%%%%%%%%%%%%
%
% Optional Glossary or Notation section, goes here
%
%%%%%%%%%%%%%%
% Glossary is only allowed in Reviews of Geophysics
% \section*{Glossary}
% \paragraph{Term}
% Term Definition here
%
%%%%%%%%%%%%%%
% Notation -- End each entry with a period.
% \begin{notation}
% Term & definition.\\
% Second term & second definition.\\
% \end{notation}
%%%%%%%%%%%%%%%%%%%%%%%%%%%%%%%%%%%%%%%%%%%%%%%%%%%%%%%%%%%%%%%%
%
%  ACKNOWLEDGMENTS

\begin{acknowledgments}
ABS was supported by an NWO-Valorization project at CWI and by STW-project 10118. JT was supported by STW-project 10755.
\end{acknowledgments}

\end{article}

 \begin{figure}
 \begin{center}
 \noindent\includegraphics[width=0.45\textwidth]{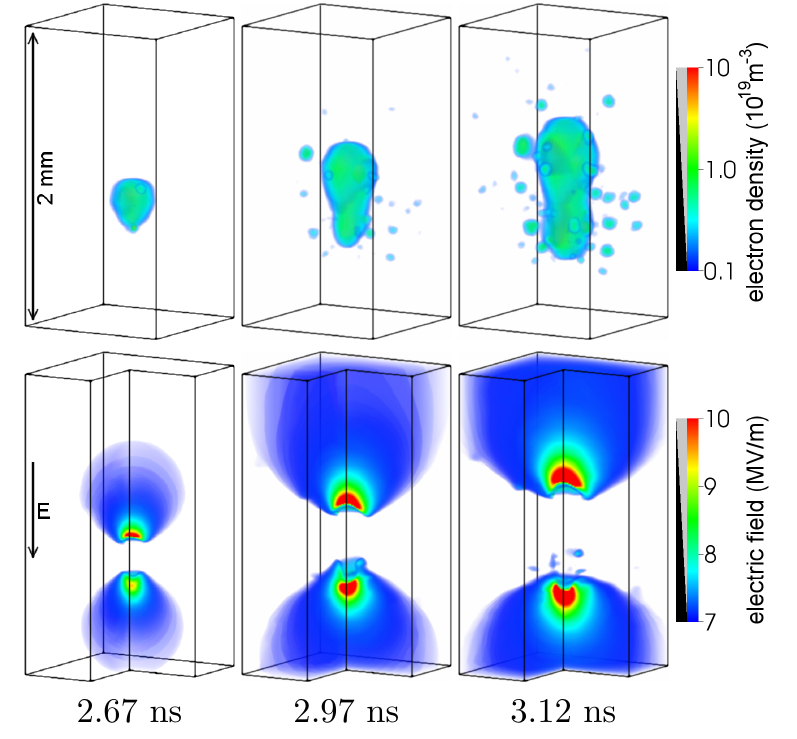}
 \caption{The electron density (top row) and the electric field (bottom row) using photoionization only (unrealistic).
 Times are indicated below each column.
 The simulation started with a single electron-ion pair in non-ionized air at 1 bar and 293 K in a downward homogeneous background field of 7 MV/m (about 2.3 times $E_k$).
 Of the total simulation domain of (4~mm)$^3$, the range from 2 to 4 mm is shown in the vertical direction, and the range from 1.5 to 2.5 mm in the two lateral directions.
 The figures were generated using volume rendering, and the opacity is shown next to the colorbar; black indicates transparency.
 For figures in the second row, a quarter of the domain is removed to show the inner structure of electric field.
 }
 \end{center}
 \label{fig:nobgion}
 \end{figure}

 \begin{figure}
 \begin{center}
 \noindent\includegraphics[width=0.46\textwidth]{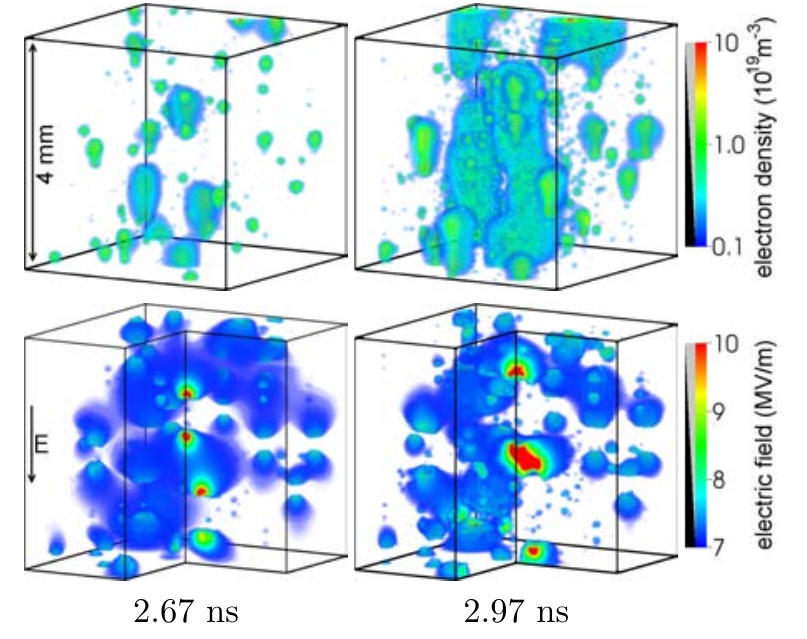}
 \caption{The electron density (top row) and the electric field (bottom row) using photoionization and natural background ionization.
 Times are indicated below each column.
 The simulation and plots were set up in the same way as for Figure~\ref{fig:nobgion},
     but now background ionization in the form of O$_2^-$ and positive ions was included, both with a density of $10^{3} \mathrm{cm}^{-3}$.
 Here the full simulation domain is shown from 0 and 4 mm in all directions.
 }
 \label{fig:bgion}
 \end{center}
 \end{figure}
 \begin{figure}
 \begin{center}
 \noindent\includegraphics[width=0.46\textwidth]{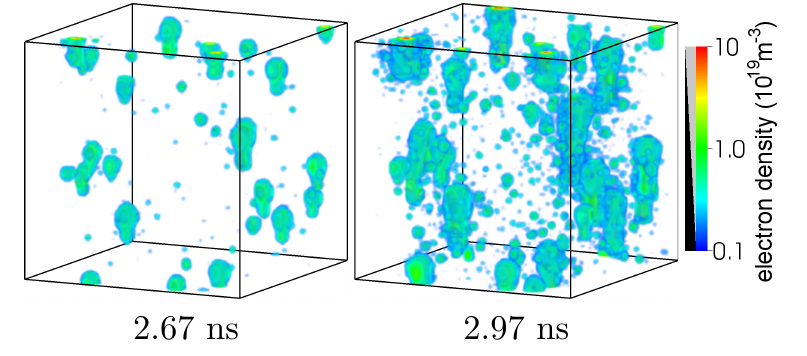}
 \caption{The electron density at 2.67 ns and 2.97 ns, using the same simulation parameters as for Figure~\ref{fig:bgion}, but now without the initial electron-ion pair.
  }
 \label{fig:fromOnlyBgion}
 \end{center}
 \end{figure}
 
 \begin{figure}
     \begin{center}
        \noindent\includegraphics[width=0.49\textwidth]{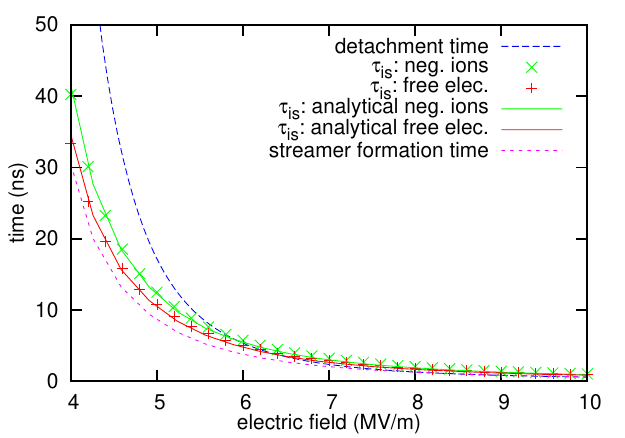}
        \caption{The ionization screening time $\tau_\mathrm{is}$ for a preionization density $n_0 = 10^3\;\mathrm{cm}^{-3}$ of electrons or negative $\mathrm{O}_2^{-}$ ions.
        The corresponding analytical approximations are also shown, see section~\ref{sec:shieldingtime}.
        Furthermore we include the detachment time and the streamer formation time, based on the Raether-Meek criterion: $18/(\alpha v_d)$ at 1 bar.}
        \label{fig:shielding_time}
     \end{center}
 \end{figure}

\end{document}